\documentclass[11pt,hidelinks]{article}
\usepackage{graphicx}
\usepackage{float}
\usepackage{enumerate}
\usepackage{natbib}
\usepackage{colortbl}
\usepackage{color}
\usepackage[dvipsnames]{xcolor}
\usepackage{easyReview}

\usepackage{amsmath,amssymb,amsthm,authblk,booktabs,bm,geometry,multirow,threeparttable,multicol,ragged2e,tikz}  
\usetikzlibrary{arrows.meta}
\usepackage{derivative,esdiff}
\allowdisplaybreaks
\usepackage{hyperref,url}

\newtheorem{theorem}{Theorem}
\newtheorem{proposition}{Proposition}
\newtheorem{assumption}{Assumption}
\newtheorem{example}{Example}

\newtheorem{algorithm}{Algorithm}

\def\PN{\textsc{pn}}
\def\RR{\textsc{rr}}
\def\ACE{\textsc{ace}}
\def\AF{\textsc{af}}
\def\CI{\textsc{CI}}
\def\PT{{\rm PT}}
\def\bk{{\overline k}}
\def\Pr{\text{\rm pr}}
\def\pr{\text{\rm pr}}

\def\EIF{{\rm EIF}}
\def\Rem{{\rm rem}}
\def\expit{\text{\rm expit }}
\def\t{{ \mathrm{\scriptscriptstyle T} }}

\makeatletter
\newcommand*{\ind}{%
	\mathbin{%
		\mathpalette{\@ind}{}%
	}%
}
\newcommand*{\nind}{%
	\mathbin{
		\mathpalette{\@ind}{\not}
	}%
}
\newcommand*{\@ind}[2]{%
	\sbox0{$#1\perp\m@th$}
	\sbox2{$#1=$}
	\sbox4{$#1\vcenter{}$}
	\rlap{\copy0}
	\dimen@=\dimexpr\ht2-\ht4-.2pt\relax
	\kern\dimen@
	{#2}%
	\kern\dimen@
	\copy0 
} 
\makeatother

\def\spacingset#1{\renewcommand{\baselinestretch}%
{#1}\small\normalsize} \spacingset{1}

\title{\bf Statistical inference for the probability of necessity for causal attribution}

\author[1]{Ping Zhang}
\author[2]{Ruoyu Wang}
\author[1]{Wang Miao}
  
\affil[1]{Department of Probability and Statistics, Peking University}
\affil[2]{Department of Biostatistics, Harvard University}
\date{}

\begin{document}
\maketitle
\begin{abstract}
To answer questions of ``causes of effects'', the probability of necessity was previously introduced for assessing whether an observed outcome was caused by an earlier treatment. 
However, statistical inference for the probability of necessity is understudied due to several difficulties, which hinder its application in practice. 
The evaluation of the probability of necessity involves the joint distribution of potential outcomes, and thus it is generally not point identified and one can at best obtain lower and upper bounds even in randomized experiments, unless fairly stringent monotonicity assumptions on potential outcomes are made. 
Moreover, these bounds are non-smooth functionals of the observed data distribution and standard estimation and inference methods cannot be directly applied.
In this paper, we investigate the statistical inference for the probability of necessity in general situations where it may not be point identified.
We introduce a mild margin condition to tackle the non-smoothness, under which the bounds become pathwise differentiable.
We establish the semiparametric efficiency theory and propose novel asymptotically efficient estimators of the lower and upper bounds, and further construct confidence intervals for the probability of necessity based on the proposed bound estimators.
The resultant confidence intervals can effectively utilize the observed covariates to reduce lengths.
The proposed approach has potential application in biomedical, epidemiological, and legal studies where understanding causal attribution beyond traditional causal effects is essential.
\end{abstract}

\noindent%
\vfill

\spacingset{1.5} 

\section{Introduction}

Causal attribution assesses how likely an observed outcome was caused by a prior treatment,
which is a different problem from evaluating the effect of a treatment on future outcomes.
The former concerns causes of effects while the latter concerns effects of causes \citep{dawid2014fitting,pearl2015causes,dawid2022effects}.
Investigating causes of effects is crucial to understanding causal mechanisms, with applications arising in epidemiology, biomedicine, and legal studies \citep{lagakos1986assigned, robins1989probability, greenland2000epidemiology, pearl1999probabilities, tian2000probabilities, dawid2014fitting}. 
A prominent attribution measure is the probability of necessity ($\PN$), also referred to as probability of causation, defined as the probability that the outcome would not have occurred without the cause, given that the outcome has occurred under the cause \citep{pearl1999probabilities,dawid2017probability}. 
Unlike standard causal effect estimands such as average causal effect, $\PN$ involves the joint distribution of potential outcomes, so conventional methods for effects of causes are not directly applicable.
Recent work has extended the concept of $\PN$ to causal attribution analysis with multiple causes or outcomes \citep{lu2023evaluating,li2024retrospective}.

Although conceptually intriguing, causal attribution measures have not been widely applied in practice,
largely due to the lack of reliable statistical inference methods.
Existing approaches \citep{cai2005variance,kuroki2011statistical,cuellar2020non,tian2025semiparametric} typically rely on point identification of $\PN$, which requires the no unmeasured confounding assumption and the monotonicity of potential outcomes in the treatment \citep{pearl1999probabilities, tian2000probabilities}.
While the no unmeasured confounding assumption is plausible in certain situations, 
monotonicity may fail even in randomized experiments, for example,
when the treatment benefits most of the population but is less effective than the control in certain subgroups.
Without the monotonicity assumption, $\PN$ is in general not point identified
and one can at best obtain lower and upper bounds.
\citet{pearl1999probabilities} and \citet{tian2000probabilities}
established bounds for  $\PN$ when solely the treatment and outcome are available, 
while \citet{dawid2017probability} proposed  narrower bounds by incorporating additional covariates.
However, inference for $\PN$ is challenging because the bounds are  non-smooth functionals of the observed data distribution,
and their estimation and uncertainty quantification are not entirely clear in the literature.
Moreover, confidence intervals for $\PN$ itself differ from confidence intervals for bounds,  
as the latter are  for the entire identified set that encompasses all possible values of $\PN$, 
which is in general overly conservative for $\PN$ itself.

In this paper, we develop a framework for inference on $\PN$ in randomized experiments without monotonicity, where $\PN$ is generally only partially identified.
We show that the covariate-assisted bound of \citet{dawid2017probability} is sharp without additional assumptions.
Because the bound functionals involve maximization operators, they are non-smooth and do not admit regular estimation without further conditions \citep{hirano2012impossibility}.
To address this challenge, we introduce a mild margin condition under which the bounds become pathwise differentiable, derive efficient influence functions, and construct asymptotically efficient estimators.
The resulting confidence intervals for $\PN$, based on asymptotically efficient estimators of the sharp bounds,
can effectively utilize covariates to shorten lengths.
We illustrate the proposed method through simulations and an application to the Licorice Gargle Data \citep{ruetzler2013randomized}. 
Possible extensions to observational studies and other causal attribution measures are discussed at the end of the paper.

\section{Causal attribution measure and the challenge for  inference}\label{sec:pre}
\subsection{Probability of necessity  for causal attribution}

Consider a randomized experiment with   a sample of  $N$ units  from a superpopulation.
Let $X$ denote  a binary treatment,  $Y$ a binary outcome, and $V$ possibly a vector of covariates.
For each unit $i = 1,\ldots,N$, 
let $X_i=1$ and $X_i=0$ denote the presence and absence of the treatment, respectively,
and $Y_i=1$ and $Y_i=0$ denote   the occurrence and  not occurrence of the effect, respectively.
Let $Y_i(x)$ denote the  potential outcome had unit $i$  received treatment $X_i=x$ for $x\in \{0,1\}$. 
The observed outcome is assumed to be a realization of the potential outcome under the treatment actually received, i.e.,  
$Y_i = X_iY_i(1) + (1-X_i)Y_i(0)$.

Most causal inference studies focus on causal effects, such as the average causal effect $\ACE=E\{Y(1)-Y(0)\}$ and the relative risk $\RR = \Pr\{Y(1)=1\}/\Pr\{Y(0)=1\}$.
However, in many applications, researchers are interested in 
causal attribution, i.e., how likely a given effect $Y = 1$ is due to a suspected cause $X = 1$.
A conventional attribution measure in epidemiology is the attributable fraction among the exposed \citep{rothman2008modern},
$\AF = 1 - \pr\{Y(0)=1\mid X=1\}/\pr(Y=1\mid X=1)$,
which can be viewed as a normalized causal effect and requires additional assumptions for a causal attribution interpretation \citep{pearl1999probabilities}.
Based on counterfactual reasoning,
\citet{pearl1999probabilities} introduced the probability of necessity for causal attribution,
also referred to as probability of causation \citep{dawid2017probability}:
\[\PN = \Pr\{Y(0)=0 \mid X=1,Y=1\}.\]
This parameter, defined in the subpopulation who have received the  treatment and experienced the effect (i.e. $X=1, Y=1$), 
assesses the probability of the effect not occurring had   the subpopulation not been exposed to the treatment (i.e. $Y(0) = 0$).
It can be  equivalently written as $\PN = E\{Y(1)-Y(0)\mid X=1,Y=1\}$, 
which differs from traditional  causal effect estimands because of conditioning on $Y$.
It can also be expressed as $\PN = 1 - \pr\{Y(0)=1,Y(1)=1\mid X=1\}/\pr(Y=1\mid X=1)$,
implying that $\PN \geq \AF$.

For conceptual illustration of $\PN$, suppose Alice took a drug for $3$ years and developed diabetes, 
and given this fact she asked: What would have happened had she not taken the drug? 
This is a question about the attribution of an effect that has already occurred,
which is crucial for clinicians and epidemiologists to evaluate whether diabetes of Alice is a result of taking the drug.
It is different from the question about the effects of applied causes in traditional causal inference: What would happen to Alice if she were to take the drug? 
The former question of causal attribution is generally useful for understanding the effect occurred in the past, 
while causal effect estimands  are most useful for guiding future interventions \citep{yamamoto2012understanding}.
The concept of $\PN$ and its   application are appealing across various fields.
It can be used to determine the likelihood that a patient's cancer was caused by radiation and make compensation policy in legal cases \citep{lagakos1986assigned}.
In particular, $\PN \geq 50\%$ is often interpreted as ``more probable than not'' \citep{pearl1999probabilities, dawid2014fitting},
which is an important indicator for  whether the risk factor at issue is more likely than not responsible for a patient's disease,
and some courts invoke this as important evidence.
For a real legal case example concerning attribution of cancer, \citet{silicone_gel_case}, handled by the United States District Court for the Central District of California, requires ``doubling of the risk'' (i.e., $\RR \geq 2$) to prove specific causation (e.g., the plaintiff's breast cancer was caused by her implants).
This requirement arises because $\RR \geq 2$ implies $\PN \geq 50\%$ in the absence of confounding.

We make the following assumption for  randomized experiments.
\begin{assumption}\label{ass:re}
(i) Randomized experiment: $X \ind \{Y(0),Y(1),V\}$;

(ii) Positivity: $0 < \Pr(X=1) <1$;

(iii) $\Pr\{Y(1)=1\} >0$ and $0 < \Pr\{Y(0) = 1\} < 1$.
\end{assumption}
In addition to the random assignment of treatment described in condition (i), 
condition (ii) requires positive probability of  both the treatment and control assignment,
which is standard in causal inference.
Condition (iii)  guarantees that $\Pr(X=1,Y=1) >0$ so that  $\PN$ is well defined and excludes the trivial cases in which the potential outcomes have no randomness.

Under Assumption \ref{ass:re}, $\ACE$, $\RR$, and $\AF$ are identified from the observed data.
However, $\PN = 1-\Pr\{Y(0) = 1, Y(1) = 1\} / \Pr\{Y(1) = 1\}$ depends on the joint distribution of potential outcomes and is generally not identified.
Under the monotonicity assumption  $Y(1) \geq Y(0)$, 
the joint distribution of potential outcomes can be represented by the marginal distributions, $\Pr\{Y(0) = 1, Y(1) = 1\} =  \Pr\{Y(0) = 1\}$, 
so that $\PN = \AF = 1 - 1/\RR$.
In randomized experiments with the monotonicity assumption, 
causal attribution based on $\PN$ is equivalent to using $\AF$ or $\RR$ by noting that  $\PN \geq 50\%$ if and only if $\RR \geq 2$.
However, this equivalence relies critically on monotonicity,
which is fairly stringent in practice.
Throughout this paper, we focus on the statistical inference for $\PN$  without the monotonicity assumption.
In this setting, $\PN$ is no longer equal to $\AF$;
instead, it can only be partially identified.

\begin{proposition}\label{prop:bounds}
Under Assumption \ref{ass:re},  
(i) without additional covariates,  \citet{pearl1999probabilities} and \citet{tian2000probabilities} obtained that $L_\PT \leq \PN \leq U_{\PT}$,
\begin{eqnarray*}
L_{\PT} & = & \max\left\{0, \frac{\Pr(Y=1\mid X=1) - \Pr(Y=1 \mid X=0)}{\Pr(Y=1\mid X=1)}\right\},\\
U_{\PT} & = & 1 - \max \left\{0, \frac{\Pr(Y=1\mid X=1) - \Pr(Y=0\mid X=0)}{\Pr(Y=1 \mid X=1)}\right\};
\end{eqnarray*}
(ii) with the assist of covariates $V$, 
\citet{dawid2017probability} obtained that $L \leq \PN \leq U$,
\begin{eqnarray*}
L &=& E\left[ \max\left\{0, \frac{\Pr(Y=1\mid X=1,V) - \Pr(Y=1\mid X=0,V)}{\Pr(Y=1\mid X=1)}\right\} \right], \\
U &=& E\left[ 1 - \max\left\{0, \frac{\Pr(Y=1\mid X=1,V) - \Pr(Y=0\mid X=0,V)}{\Pr(Y=1\mid X=1)}\right\} \right] .
\end{eqnarray*}
\end{proposition}
These bounds characterize possible values of $\PN$ compatible with the observed data. Analogous bounding approaches have also been developed in the principal stratification literature, where parameters of interest are likewise defined based on joint potential outcomes; see, for example, the work on truncation by death \citep{zhang2003estimation, grilli2008nonparametric}.
\citet{tian2000probabilities} showed that the no-covariate bound is sharp in randomized experiments without covariates,
in the sense that for any observed data distribution $\pr(X,Y)$, there exists some joint distribution $\pr\{X,Y(0),Y(1)\}$ that is consistent with $\pr(X,Y)$, such that $\PN$ can be any value permitted by the bound.
With additional covariates, \citet{dawid2017probability} pointed out that the covariate-assisted bound is  tighter  than the no-covariate bound because $L_{\PT} \leq L$ and $U_{\PT} \geq U$. 
We  prove that the covariate-assisted bound is sharp.
\begin{proposition}\label{prop:sharp}
Under Assumption \ref{ass:re}, the bound $\PN \in [L,U]$ is sharp, 
in the sense that for any observed data distribution $\pr(X,Y,V)$, there exists some $\pr\{X,Y(0),Y(1),V\}$ that is consistent with $\pr(X,Y,V)$, such that $\PN$ can be  any value between $L$ and $U$.
\end{proposition}
The covariate-assisted bound cannot be further improved without additional assumptions.
Moreover, the lower bounds $L_{\PT}$ and $L$ imply that $\PN \geq \AF=1-1/\RR$,
so $\PN \geq 50\%$ whenever  $\RR \geq 2$.
However,  when monotonicity fails,   
$\PN$ may exceed $50\%$ even when $\RR < 2$ \citep{greenland1999relation,dawid2014fitting}, i.e. the causal effect is small.
The following is an example illustrating that  $\PN$ could exceed $50\%$ while $\RR <1$ and $\AF<0$.
\begin{example}
Suppose  $V$ is a binary covariate  with $\Pr(V=1) = \Pr(V=0) = 0.5$.
The conditional probabilities are specified as follows: $\Pr(Y = 1\mid X=1,V=1) = 0.4$, $\Pr(Y = 1\mid X=1,V=0) = 0.1$, $\Pr(Y = 1\mid X=0,V=1) = 0.1$, and $\Pr(Y = 1\mid X=0,V=0) = 0.8$.
Then under Assumption \ref{ass:re}, $\ACE=-0.2$,  $\RR =5/9$, $\AF=-0.8$, and $L_\PT = 0$, 
but in contrast, we can conclude that $\PN \geq L = 0.6$.
\end{example}
This example suggests that $\PN$ can differ substantially from $\AF$.
Therefore, conventional estimands such as $\AF$ and $\RR$ are insufficient for causal attribution.
It also demonstrates that the covariate-assisted bound $(L,U)$ can  be much more informative for $\PN$ than the no-covariate bound.
Although the bounds in Proposition \ref{prop:bounds} are available in the literature,
statistical inference for $\PN$ is not well studied when it is not point identified.
We aim to construct confidence intervals for $\PN$ based on these sharp bounds.

\subsection{Challenge for  inference}\label{sec:ci}

In randomized experiments with the monotonicity assumption, $\PN$ is point identified and standard Wald-type confidence intervals can be constructed. 
Let $\widehat\PN$ and $\widehat\sigma^2$ denote estimators of $\PN$ and its asymptotic variance. 
A confidence interval for $\PN$ is
\begin{eqnarray*}
\CI_1 &=& [\widehat\PN - \Phi^{-1}(1-\alpha/2)\widehat\sigma N^{-1/2},\widehat\PN + \Phi^{-1}(1-\alpha/2)\widehat\sigma N^{-1/2}] \cap [0,1],
\end{eqnarray*}
where $\Phi(\cdot)$ is the standard normal distribution function.
However, without the monotonicity assumption,
$\PN$ is generally not point identified and $\CI_1$ is no longer valid.

Without monotonicity, confidence intervals for $\PN$ crucially rely on the lower and upper bounds.
Suppose that $N^{1/2}$-consistent and asymptotically normal estimators $\widehat L$ and $\widehat U$ of $L$ and $U$ are available, with variance estimators $\widehat\sigma_L^2$ and $\widehat\sigma_U^2$. 
A natural confidence interval is
\begin{eqnarray*}
\CI_2&=&  [\widehat{L} - \Phi^{-1}(1-\alpha/2)  \widehat{\sigma}_L N^{-1/2}, \  \widehat{U} + \Phi^{-1}(1-\alpha/2)  \widehat{\sigma}_U N^{-1/2} ] \cap [0,1],
\end{eqnarray*}
which covers the identified set determined by the bounds with asymptotical confidence level $1-\alpha$,
i.e., $\lim \inf_{N \rightarrow \infty} \Pr([L,U] \subseteq \CI_2)  \geq 1-\alpha$,
and thus $\lim \inf_{N \rightarrow \infty} \Pr(\PN\in \CI_2)  \geq 1-\alpha$.
However, $\CI_2$ is generally conservative for $\PN$ because it targets the entire identified set $[L,U]$ rather than $\PN$ itself. 
Following \citet{imbens2004confidence}, we instead consider
\begin{eqnarray*}
\CI_3 & = & [\widehat{L} - C  \widehat{\sigma}_L N^{-1/2}, \  \widehat{U} + C  \widehat{\sigma}_U N^{-1/2} ] \cap [0,1],
\end{eqnarray*}
where the critical value $C$ is the unique solution to
\begin{eqnarray*}
\Phi \left[ C + \frac{N^{1/2}(\widehat{U} - \widehat{L}) }{ \max\{ \widehat{\sigma}_L, \widehat{\sigma}_U \} } \right] - \Phi(-C) = 1-\alpha.
\end{eqnarray*}
According to \citet{imbens2004confidence}, 
$\CI_3$ has asymptotical confidence level $1-\alpha$,
i.e., $\lim \inf_{N \rightarrow \infty} \Pr(\PN\in \CI_3)  \geq 1-\alpha$.
Moreover, $\CI_3$ is generally narrower than $\CI_2$ because $C$ lies between $\Phi^{-1}(1-\alpha)$ and $\Phi^{-1}(1-\alpha/2)$ when $\widehat U \geq \widehat L$, and is thus preferred in practice.
One-sided confidence intervals for hypothesis testing are described in Web~Appendix~A.

Nonetheless, these confidence intervals entail   $N^{1/2}$-consistent and asymptotically normal estimators of $L$ and $U$.
However, the bounds are non-smooth functionals of the observed data distribution, and their estimation is challenging.
Standard plug-in estimators are generally not asymptotically normal. 
In addition, efficient estimation of the bounds is desirable for constructing narrower confidence intervals, but the corresponding semiparametric efficiency theory is not clear.
In the following, we propose a feasible approach for efficiently estimating the bounds, which paves the way for constructing confidence intervals for $\PN$.

\section{Statistical inference without the monotonicity assumption}\label{sec:cov}
We consider estimating the covariate-assisted bound $(L,U)$ and constructing confidence intervals for $\PN$.
Let $p = \pr(X=1)$,
$\mu_{yx} = \pr(Y=y\mid X=x)$,
and $\pi(X,V)$ $= \Pr(Y=1\mid X, V)$.
Define $d_L(V) = 1\{\pi(1,V) > \pi(0,V) \}$,  $d_U(V) = 1\{\pi(1,V) > 1-\pi(0,V) \}$,
and $d(V) = \{d_L(V), d_U(V)\}$.
The lower and upper bounds  can be equivalently written as
\[
L =  \frac{\Delta_L}{\mu_{11}}, \quad
U = 1 - \frac{\Delta_U}{\mu_{11}},
\]
with $\Delta_L = E[d_L(V)\{\pi(1,V)-\pi(0,V)\}]$ and $\Delta_U = E[d_U(V)\{\pi(1,V)+\pi(0,V)-1\}]$.
Without additional assumptions, the covariate distribution may place substantial probability mass near the non-differentiability points.
In this case, the estimators may suffer from non-negligible bias, and regular estimation is generally unattainable \citep{hirano2012impossibility}.
We introduce the following assumption to address the non-smoothness.
\begin{assumption}[Margin condition] \label{ass:margin}
There exist constants $\gamma > 0$ and $c>0$ such that 
$\Pr \{|\pi(1,V) - \pi(0,V)| \leq t\} \leq c t^\gamma$ and
$\Pr \{|\pi(1,V) + \pi(0,V)-1| \leq t\} \leq c t^\gamma$ for any $t>0$.
\end{assumption}

Assumption~\ref{ass:margin} imposes a local rate restriction on the probability mass of the contrast functions $\pi(1,V)-\pi(0,V)$ and $\pi(1,V)+\pi(0,V)-1$ in neighborhoods near zero, ruling out excessive concentration at the non-differentiability points.
An intuitive sufficient condition is that the conditional average causal effect remains uniformly separated from zero across all covariate strata, and that the outcome $Y=1$ is relatively rare so that both $\pi(1,V)$ and $\pi(0,V)$ are well below $0.5$.
Under this condition, Assumption~\ref{ass:margin} holds for any $\gamma>0$.
More generally, if the densities of $\pi(1,V)-\pi(0,V)$ and $\pi(1,V)+\pi(0,V)-1$ are bounded in a neighborhood of zero, then Assumption~\ref{ass:margin} holds with $\gamma=1$ \citep{levis2025covariate}.
It is useful to distinguish Assumption~\ref{ass:margin} from monotonicity.
Monotonicity is an individual-level restriction on potential outcomes, whereas Assumption~\ref{ass:margin} is a distributional condition on the observed data functions $\pi(1,V)$ and $\pi(0,V)$.
The two assumptions play distinct roles: monotonicity is a causal structural restriction for identifying $\PN$, whereas Assumption~\ref{ass:margin} is a technical regularity condition for estimating non-smooth bounds.
Analogous conditions originate in classification \citep{audibert2007fast} and have recently been adopted in causal inference for inference on partially identified or nonregular parameters \citep{luedtke2016statistical, kennedy2019survivor, levis2025covariate, ben2025partial}.

Although Assumption~\ref{ass:margin} depends only on the observed data distribution, it is difficult to test in finite samples; see Web~Appendix~B for further discussion.
Its plausibility can be assessed using subject-matter knowledge.
Because the condition rules out substantial probability mass at covariate strata with null causal effects, it is more plausible when effects are expected to be meaningfully non-zero; for example, when the treatment and control operate through substantially different biological mechanisms.
The plausibility of Assumption~\ref{ass:margin} also depends on the level of covariate stratification.
If overly fine covariate strata lead to many near-zero contrasts, one may use a coarser, scientifically meaningful summary of the covariates.
Such coarsening can reduce concentration of the contrast functions near zero, making the margin condition more plausible, but at the cost of discarding covariate information and potentially widening the bounds.
Let $O = (X,Y,V)$ denote the observed data.
We next establish the semiparametric efficiency theory for estimating the bounds \citep{bickel1993efficient, tsiatis2006semiparametric}.

\begin{theorem} \label{thm:eif}
Under Assumptions \ref{ass:re} and \ref{ass:margin}, $L$ and $U$ are pathwise differentiable and the efficient influence functions for $L$ and $U$ are
\[\EIF_L(O;\eta,d)
= \psi_L(O;\eta,d) - L,\quad
\EIF_U(O;\eta,d)
= \psi_U(O;\eta,d) - U,\]
where $\eta = (\mu_{11},p, \Delta_L, \Delta_U,\pi)$ denotes  the nuisance parameters and functions,
and 
\begin{eqnarray*}
\psi_L(O;\eta,d) & = & \frac{d_L(V)}{\mu_{11}}  \left[ \pi(1,V) - \pi(0,V) + \frac{X\{Y-\pi(1,V)\}}{p} - \frac{(1-X)\{Y-\pi(0,V)\}}{1-p} \right]\nonumber\\
&&- \frac{\Delta_L}{\mu_{11}^2} \left[ \frac{X\{Y - \pi(1,V)\}}{p} + \pi(1,V) - \mu_{11} \right],\\
\psi_U(O;\eta,d) & = & 1 - \frac{d_U(V)}{\mu_{11}} \left[ \pi(1,V) + \pi(0,V)-1 + \frac{X\{Y-\pi(1,V)\}}{p} + \frac{(1-X)\{Y-\pi(0,V)\}}{1-p} \right]\nonumber\\
&& + \frac{\Delta_U}{\mu_{11}^2} \left[\frac{X\{Y - \pi(1,V)\}}{p} + \pi(1,V) -\mu_{11} \right].
\end{eqnarray*}
\end{theorem}
Theorem \ref{thm:eif} provides the closed-form efficient influence functions for bound functionals.
Under Assumptions \ref{ass:re} and \ref{ass:margin}, the semiparametric efficiency bounds for estimating $L$ and $U$ are $E\{\EIF_L(O;\eta,d)^2\}$ and $E\{\EIF_U(O;\eta,d)^2\}$, respectively.
The semiparametric efficiency bounds of $L$ and $U$ represent the best possible precision achievable by any regular and asymptotically linear estimators under our assumptions.
Efficient estimation of $L$ and $U$ is therefore desirable for constructing confidence intervals for $\PN$.

Motivated by Theorem \ref{thm:eif},
we estimate $L$ and $U$ by first estimating the nuisance parameters and then plugging in them into efficient influence functions.
We use cross-fitting \citep{chernozhukov2018double}, which allows flexible nuisance estimators outside Donsker classes \citep{van1996weak}.
Algorithm~\ref{algo} summarizes the procedure for estimation.

\begin{algorithm}\label{algo}
the proposed bound estimators

Input: A sample of $N$ observations, $(X_i,Y_i,V_i)_{i=1}^N$.
	
Step 1: randomly split the sample into $K$ folds indexed by $I_1,\ldots,I_K$ with equal size $n=N/K$;
let $I_\bk = \{1,\ldots,N\}\backslash I_k$ denote the remaining parts.
	
Step 2: use the sample $I_\bk$ to estimate nuisance parameters.
Calculate the quantities $\widetilde{\eta}^{(k)} = (\widetilde\mu_{11}^{(k)}, \widetilde p^{(k)},\widetilde\Delta_{L}^{(k)}, \widetilde\Delta_{U}^{(k)},\widetilde\pi^{(k)})$ and  $\widetilde d^{(k)}(V) = \{\widetilde d_L^{(k)}(V),\widetilde d_U^{(k)}(V)\}$ with $\widetilde d_{L}^{(k)}(V) = 1\{\widetilde\pi^{(k)}(1,V) > \widetilde\pi^{(k)}(0,V)\}$ and $\widetilde d_U^{(k)}(V) = 1\{\widetilde\pi^{(k)}(1,V) > 1-\widetilde\pi^{(k)}(0,V)\}$ being the nuisance estimators.
	
Step 3: obtain $\widehat{L}^{(k)} = \widehat E_k\{\psi_L(O; \widetilde{\eta}^{(k)}, \widetilde{d}^{(k)})\}$ and $\widehat{U}^{(k)} = \widehat E_k\{\psi_U(O; \widetilde{\eta}^{(k)}, \widetilde{d}^{(k)})\}$,
where $\widehat E_k$ is the empirical expectation over sample $I_k$.
	
Step 4: obtain bound estimators $\widehat{L}$ and $\widehat{U}$ by averaging over  $\widehat{L}^{(k)}$ and $\widehat{U}^{(k)}$ for all $k$,
\[\widehat{L} = \frac{1}{K} \sum_{k=1}^K \widehat{L}^{(k)},\quad
\widehat{U} = \frac{1}{K} \sum_{k=1}^K \widehat{U}^{(k)}.\]
\end{algorithm}

In the first step, we randomly split the sample  into $K$ parts with equal size. 
In the second step, we obtain estimators  $\widetilde\eta^{(k)}$ and  $\widetilde d^{(k)}(V)$ using $I_\bk$;
and then in the third step we estimate $L$ and $U$ by evaluating the efficient influence functions in Theorem \ref{thm:eif} using $I_k$.
In the last step, we aggregate the estimators for $k=1,\ldots,K$ to obtain the final  bound estimators.

We state how to estimate the nuisance parameters in the second step of Algorithm \ref{algo}.
The nuisance $\eta$ consists of three components:
the smooth parameters $p$ and $\mu_{11}$,
the function $\pi(X,V)$,
and the non-smooth parameters $\Delta_L$ and $\Delta_U$.
The smooth parameters are estimated by their sample analogues $\widetilde p^{(k)} = \widetilde E_k(X)$ and $\widetilde \mu_{11}^{(k)} = \widetilde E_k(XY) / \widetilde E_k(X)$, where $\widetilde E_k$ denotes the empirical expectation over sample $I_\bk$.
The function $\pi(X,V)$ can be estimated using standard classification methods.
For example, we can specify a parametric working model $\pi(X,V;\beta)$ and solve the score equation using $I_\bk$ to obtain $\widetilde\beta^{(k)}$ and $\widetilde\pi^{(k)}(X,V) = \pi(X,V;\widetilde\beta^{(k)})$.
More flexible semiparametric or nonparametric working models may also be used.
In the simulations and real data application, we implement logistic regression and Super Learner, an ensemble approach combining multiple user-specified algorithms \citep{polley2019package}.
The non-smooth parameters are estimated using plug-in estimators $\widetilde \Delta_{L}^{(k)} = \widetilde E_k[\widetilde d_L^{(k)}(V) \{ \widetilde \pi^{(k)}(1,V) - \widetilde \pi^{(k)}(0,V)\}]$ and $\widetilde \Delta_U^{(k)} = \widetilde E_k[\widetilde d_{U}^{(k)}(V) \{ \widetilde \pi^{(k)}(1,V) + \widetilde \pi^{(k)}(0,V)-1\}]$.

The following theorem characterizes the asymptotic distribution of $(\widehat L, \widehat U)$.
We use the $L_\infty$ norm to measure the convergence rate of the nuisance estimator, $||\widetilde\pi^{(k)}-\pi||_\infty = \inf\{M\geq 0: |\widetilde\pi^{(k)}(X,V)-\pi(X,V)| \leq M \text{ almost surely}\}$.
\begin{theorem}\label{thm:asymp}
Under Assumptions \ref{ass:re} and \ref{ass:margin}, suppose the nuisance estimators satisfy that
(i) $\widetilde p^{(k)}$ and $\widetilde \mu_{11}^{(k)}$ are $N^{1/2}$-consistent and asymptotically normal estimators of $p$ and $\mu_{11}$, respectively;
(ii)  $|| \widetilde\pi^{(k)} - \pi  ||_\infty = o_p(1)$;
(iii) $\widetilde \Delta_L^{(k)}$ and $\widetilde \Delta_U^{(k)}$ are consistent estimators of $\Delta_L$ and $\Delta_U$, respectively.
Then we have
\begin{eqnarray}
\widehat L - L & = & N^{-1}\sum_{i=1}^N\EIF_L(O_i;\eta,d)
+ O_p(\Rem)
+ o_p (N^{-1/2}),\label{L:expansion}\\
\widehat U - U & = &  N^{-1} \sum_{i=1}^N\EIF_U(O_i;\eta,d)
+ O_p(\Rem)
+ o_p (N^{-1/2}),\label{U:expansion}
\end{eqnarray}
where $\Rem = \max_{1\leq k\leq K} || \widetilde\pi^{(k)} - \pi  ||_\infty^{1+\gamma}.$
\end{theorem}
Theorem \ref{thm:asymp} presents the asymptotic expansion of the bound estimators under mild conditions on nuisance estimators.
Conditions (i)-(iii) in Theorem \ref{thm:asymp} can be satisfied by the estimators used in Algorithm \ref{algo}:
$\widetilde p^{(k)}$ and $\widetilde \mu_{11}^{(k)}$ satisfy Condition (i);
$\widetilde \pi^{(k)}(X,V)$ satisfies Condition (ii) if we correctly specify parametric working models or adopt nonparametic estimation methods;
Condition (iii) follows from Condition (ii) for $\widetilde \Delta_L^{(k)}$ and $\widetilde \Delta_U^{(k)}$; see Web~Appendix~D.
According to Equations \eqref{L:expansion} and \eqref{U:expansion},
$\widehat L$ and $\widehat U$ are $N^{1/2}$-consistent and asymptotically normal with influence functions $\EIF_L(O;\eta,d)$ and $\EIF_U(O;\eta,d)$, respectively,
as long as the bias term $\Rem$ is $o_p (N^{-1/2})$.
Equivalently, this requires $|| \widetilde\pi^{(k)} - \pi  ||_\infty = o_p(N^{-1/(2+2\gamma)})$.

\begin{proposition}\label{prop:eff}
Under the conditions of Theorem \ref{thm:asymp} and  that $|| \widetilde\pi^{(k)} - \pi  ||_\infty = o_p(N^{-1/(2+2\gamma)})$,  
$\widehat L$ and $\widehat U$ are asymptotically efficient.
\end{proposition}
Proposition \ref{prop:eff} requires the nuisance estimator $\widetilde \pi^{(k)}$ to converge faster than $O_p(N^{-1/(2+2\gamma)})$ for efficient estimation of $L$ and $U$.
Under correctly specified parametric working models, the convergence rate is $O_p(N^{-1/2})$, which satisfies the requirement for all $\gamma>0$;
however, $\widetilde \pi^{(k)}$ may not be consistent when models are misspecified.
For flexible or nonparametric methods, the convergence rate is generally slower than $O_p(N^{-1/2})$ and whether the condition holds depends on $\gamma$.
In particular, when Assumption \ref{ass:margin} holds with $\gamma=1$, the required rate becomes $o_p(N^{-1/4})$, which is a standard requirement for nuisance estimators in semiparametric estimation of complicated functionals.
Such rates can be achieved by a broad class of flexible estimation methods, including random forests, boosted trees, neural networks, and ensemble methods \citep{chernozhukov2018double}.
Under conditions of Proposition \ref{prop:eff},  
the asymptotic variances of $\widehat L$ and $\widehat U$ attain the semiparametric efficiency bounds $E\{\EIF_L(O;\eta,d)^2\}$ and $E\{\EIF_U(O;\eta,d)^2\}$, respectively,
which can be consistently estimated with
\[\widehat \sigma_L^2 = \frac{1}{K} \sum_{k=1}^K \widehat E_k[\{\psi_L(O;\widetilde\eta^{(k)},\widetilde d^{(k)})-\widehat L\}^2], \quad
\widehat \sigma_U^2 = \frac{1}{K} \sum_{k=1}^K \widehat E_k[\{\psi_U(O;\widetilde\eta^{(k)},\widetilde d^{(k)})-\widehat U\}^2].
\]

Therefore,
we can apply the confidence intervals for $\PN$ discussed in Section \ref{sec:ci}:
\begin{eqnarray*}
\CI_2 &=&  [\widehat{L} - \Phi^{-1}(1-\alpha/2)  \widehat{\sigma}_L N^{-1/2}, \  \widehat{U} + \Phi^{-1}(1-\alpha/2)  \widehat{\sigma}_U N^{-1/2} ] \cap [0,1],\\
\CI_3 &=& [\widehat{L} - C  \widehat{\sigma}_L N^{-1/2}, \  \widehat{U} + C  \widehat{\sigma}_U N^{-1/2} ] \cap [0,1],
\end{eqnarray*}
where the critical value $C$ is the unique solution to
\[
\Phi \left[ C + \frac{N^{1/2}(\widehat{U} - \widehat{L}) }{ \max\{ \widehat{\sigma}_L, \widehat{\sigma}_U \} } \right] - \Phi(-C) = 1-\alpha.
\]
Existing inference methods based on influence functions include \citet{cuellar2020non} and \citet{tian2025semiparametric}, which rely on stringent assumptions such as monotonicity to achieve point identification of causal attribution parameters. 
These approaches yield confidence intervals of the $\CI_1$ type, which are invalid when $\PN$ is not identified. 
In contrast, our approach does not require identification of $\PN$. 
We derive efficient influence functions for the covariate-assisted bound $(L,U)$ and construct corresponding efficient estimators, enabling valid inference for $\PN$ via $\CI_2$ and $\CI_3$ under partial identification.
Overall, we recommend using $\CI_3$ for inference on $\PN$ when covariates are available and Assumption~\ref{ass:margin} is plausible; when covariates are unavailable, we additionally provide a confidence interval based on the no-covariate bound $(L_\PT, U_\PT)$ in Web~Appendix~C.

Our inference method for $\PN$ relies on Assumption~\ref{ass:margin} for efficient estimation of the bound functionals $L$ and $U$.
When Assumption~\ref{ass:margin} is violated, the non-differentiability of the maximization operators may induce non-negligible bias, leading to overestimation of $L$ and underestimation of $U$.
Consequently, the resulting confidence intervals can become unreliable.
Simulation results in Web~Appendix~E.2 show that the coverage of the proposed method can fall below the nominal level under severe violations.
When the margin condition is doubtful, one may instead consider methods for partially identified parameters from the econometrics literature, particularly intersection bounds \citep{romano2010inference, chernozhukov2013intersection}.
However, our setting differs from the intersection-bounds framework: our bounds involve expectations of pointwise maximization operators evaluated at each covariate value, whereas existing methods mainly address bounds defined by global supremum or infimum operators.
Extending these methods to the causal attribution problem is an important direction for future research.

\section{Simulations}\label{sec:sim}

We evaluate the performance of the proposed method via simulations.
A binary treatment $X$, a binary outcome $Y$ and a vector of two covariates $V=(V_1,V_2)$ are generated from
\begin{equation*}
\begin{gathered}
V_1\ind V_2 \text{ and } V_1, V_2 \sim {\rm U}(-1,1),\\
\pi(X,V)= \expit(X -V_1 + 2V_2 + 5XV_1 -4 XV_2),\quad 
\Pr\{Y(x) = 1\mid V\} = \pi(x,V),\\
\Pr\{Y(0)=0, Y(1)=1\mid V\} =\lambda  \min\{\pi(1,V), 1-\pi(0,V)\} +(1 - \lambda)  \max\{0, \pi(1,V)- \pi(0,V)\},\\
\Pr(X=1)=0.5,\quad Y=XY(1) + (1-X) Y(0),
\end{gathered}
\end{equation*}
where $\expit(x) = \exp(x)/\{1+\exp(x)\}$ and $\lambda \in [0,1]$. 
Under this setup, the observed data distribution and the bound $(L,U)$ remain unchanged across different $\lambda$, whereas $\PN=\lambda U+(1-\lambda)L$ varies with $\lambda$.
We consider simulations with $\lambda\in\{0,0.05,0.1,0.5,0.9,0.95,1\}$.

We apply the method proposed in Section \ref{sec:cov} to estimate the bounds and construct confidence intervals for $\PN$.
To estimate $L$ and $U$,
we employ $5$-fold cross-fitting, i.e., $K = 5$,
and consider three settings with  different  working models for $\pi(X,V)$:
(i) a flexible model that separately estimates $\pi(x,V) = \Pr(Y=1\mid X=x,V)$ for $x=0,1$ using SuperLearner \citep{polley2019package}, with candidate estimation methods including simple means of outcome, generalized additive models, regression splines, regression trees, and random forests;
(ii) a  correctly specified logistic outcome model with interaction between $X$ and $V$,
\[\Pr(Y=1\mid X,V) = \expit(\beta_0 + \beta_X X+ \beta_V^\t V + \beta_{XV}^\t XV);\]
and (iii) a misspecified logistic outcome model without interaction between $X$ and $V$, 
\[\Pr(Y=1\mid X,V) = \expit(\beta_0 + \beta_X X+ \beta_V^\t V).\]
For inference, we implement four $95\%$ confidence intervals:
$\CI_1$ based on the method of \citet{cai2005variance},
$\CI_1'$ based on the efficient estimation of $\PN$ under the monotonicity assumption \citep{tian2025semiparametric},
and the proposed $\CI_2$ and $\CI_3$.

\begin{figure}[H]
\graphicspath{{Rscripts/Simulations/Results/main/}}
\centering
\includegraphics[height=0.25\textheight]{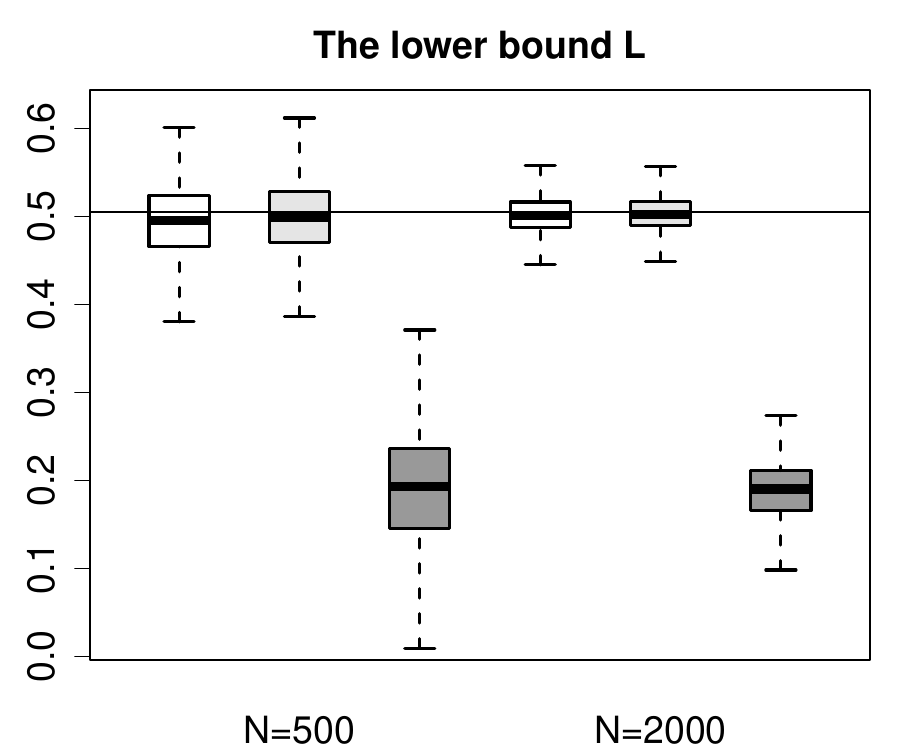}
\includegraphics[height=0.25\textheight]{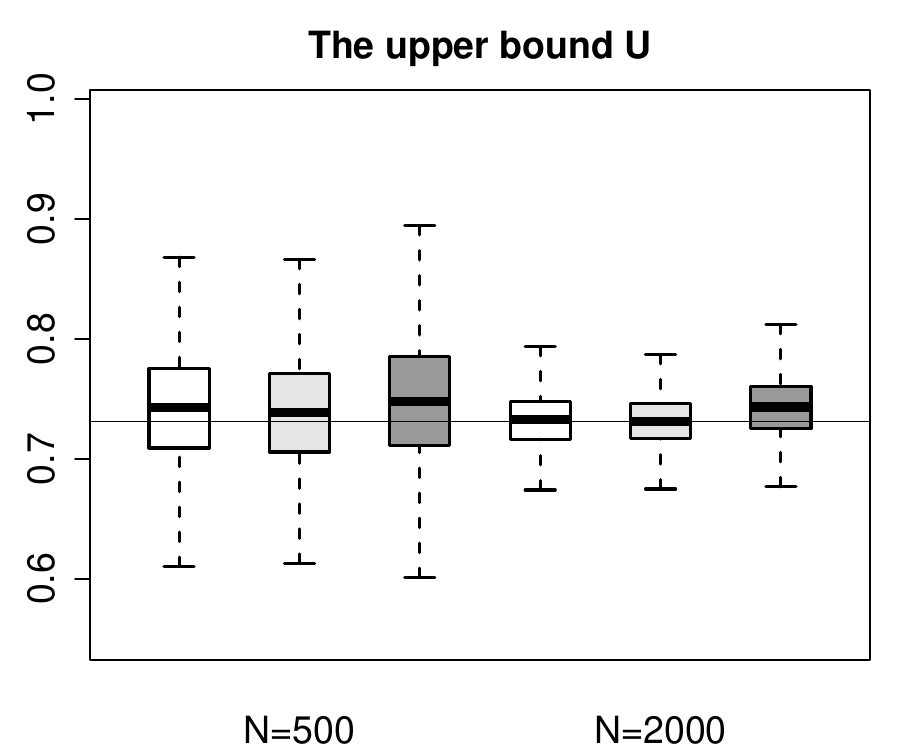}
\caption{Bias of bound estimates with different methods.}\label{fig:esti} 

\begin{justify}
Note: White boxes are for estimates in Setting (i) with flexible working model, 
light gray for Setting (ii) with correct  working model, and dark for Setting (iii) with misspecified working model. 
The horizontal lines mark the true values of $L$ and $U$, respectively.
\end{justify}
\end{figure}

For each $\lambda$, we replicate $1000$ simulations at sample sizes $N = 500$ and $2000$.
Figure~\ref{fig:esti} shows the bias of $\widehat L$ and $\widehat U$ for $\lambda = 0$.
Results for other values of $\lambda$ are relegated to Web~Appendix~E.1,
which are quite similar because the observed data distribution remains unchanged.
In Settings (i) and (ii) where the working model is flexible or  correctly specified,
the estimators have very little bias that attenuates toward zero as the sample size increases.
In Setting (iii), 
the  estimators  have   large bias, particularly for the lower bound,  due to misspecification of the working model.
Therefore, we recommend bound estimators using flexible estimation of nuisance functions, which are used in the remainder of this section.

Table~\ref{tbl:awcr} reports the average width and coverage rate of the confidence intervals.
The average widths of all confidence intervals remain largely unchanged as $\lambda$ varies, because $\lambda$ does not affect the observed data distribution and therefore does not alter the distribution of estimators.
In all settings, $\CI_1$ and $\CI_1'$ exhibit low coverage rates, whereas $\CI_2$ and $\CI_3$ maintain coverage rates close to or above the nominal level of $0.95$.
The invalidity of $\CI_1$ and $\CI_1'$ arises because the monotonicity assumption does not hold in our simulations.
Among the valid confidence intervals, $\CI_3$ consistently has the shortest width, while $\CI_2$ is wider and more conservative.
This is because $\CI_2$ is constructed to cover the entire identified set $[L,U]$, whereas $\CI_3$ targets $\PN$ directly.
For $\lambda=0$ or $1$, where $\PN$ lies on the boundary of $[L,U]$, the coverage rates of $\CI_3$ are close to the nominal level, especially for large sample sizes (e.g., $N=2000$).
For intermediate values of $\lambda$, where $\PN$ lies in the interior of $[L,U]$, the coverage rates of $\CI_3$ increase from near $0.95$ to values close to one and then decrease toward $0.95$ as $\lambda$ varies from $0$ to $1$.
This pattern reflects the intrinsic conservativeness of inference for partially identified parameters.
Because valid confidence intervals must maintain coverage over all possible values of $\PN$ within the bound $[L,U]$, inference tends to be conservative when the true parameter is well separated from the boundaries.

We also conduct additional simulations to assess robustness to violations of Assumption~\ref{ass:margin}, with results reported in Web~Appendix~E.2 due to space constraints. 
The proposed method remains stable under mild or moderate violations, although performance deteriorates under severe violations. 
In general, we recommend using $\CI_3$ to conduct inference on $\PN$ when Assumption~\ref{ass:margin} is plausible or severe violations are not expected.

\begin{table}[H]
\centering
\caption{Average width (AW) and coverage rate (CR) of 95\% confidence intervals for $\PN$}\label{tbl:awcr}
\begin{tabular}{llcccccccc}
\hline
\multicolumn{2}{c}{\multirow{2}{*}{Confidence intervals}} & \multicolumn{2}{c}{$\CI_1$} & \multicolumn{2}{c}{$\CI_1'$} & \multicolumn{2}{c}{$\CI_2$} & \multicolumn{2}{c}{$\CI_3$}\\
& & AW & CR & AW & CR & AW & CR & AW & CR  \\
\hline
\multirow{2}*{$\lambda=0$} 
& $N =500$  & 0.248 & 0.001 & 0.266 & 0.001 & 0.414 & 0.983 & 0.387 & 0.959 \\ 
& $N =2000$ & 0.128 & 0.000 & 0.136 & 0.000 & 0.313 & 0.980 & 0.300 & \textbf{0.956} \\ 
\hline
\multirow{2}*{$\lambda=0.05$} 
& $N =500$  & 0.246 & 0.000 & 0.264 & 0.000 & 0.413 & 0.991 & 0.386 & 0.982 \\ 
& $N =2000$ & 0.128 & 0.000 & 0.136 & 0.000 & 0.314 & 0.995 & 0.300 & 0.991 \\ 
\hline
\multirow{2}*{$\lambda=0.1$} 
& $N =500$  & 0.248 & 0.000 & 0.265 & 0.000 & 0.416 & 0.997 & 0.389 & 0.994 \\ 
& $N =2000$ & 0.128 & 0.000 & 0.136 & 0.000 & 0.315 & 0.999 & 0.302 & 0.998 \\ 
\hline
\multirow{2}*{$\lambda=0.5$} 
& $N =500$  & 0.248 & 0.000 & 0.266 & 0.000 & 0.413 & 1.000 & 0.386 & 1.000 \\ 
& $N =2000$ & 0.128 & 0.000 & 0.136 & 0.000 & 0.314 & 1.000 & 0.300 & 1.000 \\ 
\hline
\multirow{2}*{$\lambda=0.9$} 
& $N =500$  & 0.247 & 0.000 & 0.265 & 0.000 & 0.414 & 0.991 & 0.387 & 0.988 \\ 
& $N =2000$ & 0.128 & 0.000 & 0.137 & 0.000 & 0.314 & 0.998 & 0.301 & 0.995 \\ 
\hline
\multirow{2}*{$\lambda=0.95$} 
& $N =500$  & 0.247 & 0.000 & 0.265 & 0.000 & 0.414 & 0.991 & 0.387 & 0.980 \\ 
& $N =2000$ & 0.128 & 0.000 & 0.136 & 0.000 & 0.313 & 0.989 & 0.300 & 0.982 \\ 
\hline
\multirow{2}*{$\lambda=1$} 
& $N =500$  & 0.247 & 0.000 & 0.265 & 0.000 & 0.412 & 0.980 & 0.385 & 0.957 \\ 
& $N =2000$ & 0.128 & 0.000 & 0.136 & 0.000 & 0.315 & 0.974 & 0.302 & \textbf{0.954} \\ 
\hline
\end{tabular}
\end{table}

\section{Real data application}\label{sec:app}

We apply the proposed method to the Licorice Gargle Data  from a randomized experiment with $236$ patients \citep{ruetzler2013randomized}.
The data are available as the \texttt{licorice\_gargle} dataset in the R package \texttt{medicaldata}.
The study compared licorice versus sugar-water gargles for preventing sore throat after intubation with double-lumen endotracheal tubes.
Sore throat outcomes were assessed at time points $0.5$ hour and $1.5$ hours after arrival at the postanesthesia care unit, and $4$ hours after extubation.
Covariates include the American Society of Anesthesiologists physical status,  body mass index,  Mallampati score, and preoperative pain.
The reported relative risks of licorice versus sugar-water at $0.5$, $1.5$, and $4$ hours are $0.54$ with $95\%$ $\CI$  ($0.30$, $0.99$), $0.31$  ($0.14$, $0.68$), and $0.48$  ($0.28$, $0.83$), respectively, 
suggesting that licorice gargling reduces the risk of sore throat compared to sugar-water.

Beyond relative risks, causal attribution questions are also of interest.
Consider patients who used a sugar-water gargle and experienced sore throat.
We ask whether their pain is attributable to the higher-risk treatment, gargling with sugar-water.
This requires assessing whether they would have avoided sore throat had they instead used the lower-risk treatment, gargling with licorice.
We focus on   $N = 233$  complete cases.
Let $Y^{(j)}= 1, j=1,2,3$ denote the occurrence of sore throat pain, defined as a sore throat pain score greater than $0$  at time points $0.5$, $1.5$ and $4$, respectively.
Let $X$ denote the treatment,   $X=1$  for sugar-water gargle and $X = 0$ for licorice gargle.
The parameters of interest are $\PN_j=\Pr\{Y^{(j)}(0) =0 \mid X=1,Y^{(j)}=1\}$, $j=1,2,3$.
Regarding the plausibility of Assumption~\ref{ass:margin}, subject-matter knowledge suggests that licorice contains bioactive compounds with anti-inflammatory and soothing properties, whereas sugar water is not expected to have comparable pharmacological effects.
Therefore, it is reasonable to expect that $\pi(1,V)-\pi(0,V)$ is not close to zero for most covariate strata.
Moreover, because sore throat is relatively rare, $\pi(1,V)+\pi(0,V)$ is expected to be well below one.
These considerations are further supported by the empirical densities of the estimates of $\pi(1,V)-\pi(0,V)$ and $\pi(1,V)+\pi(0,V)-1$ in Web~Appendix~F.1.
At time points 0.5, 1.5, and 4, the empirical densities of $\widetilde\pi^{(k)}(1,V_i)-\widetilde\pi^{(k)}(0,V_i)$ vary smoothly, with no pronounced spike at zero;
the empirical densities of $\widetilde\pi^{(k)}(1,V_i)+\widetilde\pi^{(k)}(0,V_i)-1$ are concentrated away from zero, with most values lying on the negative side.
Taken together, these findings provide no compelling evidence against Assumption~\ref{ass:margin}.
We construct $95\%$ confidence intervals for $\PN_j$ using $\CI_1$, $\CI_1'$, $\CI_2$, and $\CI_3$ introduced in Section~\ref{sec:sim}.
For estimation of $L$ and $U$, we implement $4$-fold cross-fitting, i.e., $K=4$, and use SuperLearner for nuisance estimation with the same candidate methods.

In addition to $\PN$,
we also apply the proposed method to other causal attribution measures of the form $\Pr\{Y(1-x)=1-y \mid X=x,Y=y\}$ for $x\in\{0,1\}$, $y\in \{0,1\}$ by switching labels of  $X$ or $Y$ in the definition of $\PN$.
Table~\ref{tbl:app} reports the bound estimates and confidence intervals for $\PN$ and results  for other measures are relegated to Web~Appendix~F.2.
For all time points, the covariate-assisted bounds are narrower than the no-covariate bounds.
Incorporating covariates reduces the bound widths by $3.8\%$, $7.8\%$, and $9.0\%$ at time points $0.5$, $1.5$, and $4$, respectively.
Among all four confidence intervals, $\CI_1$ has the shortest width.
The validity of $\CI_1$ and $\CI_1'$ relies on the monotonicity assumption,
under which $L = L_\PT$ should hold;
however, the estimates of $L$ are larger than those of $L_\PT$ for all time points,
suggesting possible violations of monotonicity.
Among confidence intervals that do not rely on monotonicity, $\CI_3$ is consistently narrower than $\CI_2$ because it directly targets inference on $\PN$ rather than the interval $[L,U]$.
The results of $\CI_3$ suggest that, 
among patients who gargled with sugar water and experienced sore throat pain at $0.5$, $1.5$, and $4$ hours, respectively,
at least $29.5\%$, $57.8\%$, and $42.2\%$ would not have experienced this outcome had they instead gargled with licorice, with $95\%$ confidence.
For $Y^{(2)}$ at time point $1.5$, both $\CI_2$ and $\CI_3$ exclude $0.5$,
suggesting that gargling with sugar-water instead of licorice is more likely than not responsible for the sore throat pain.
In other words, among patients who gargled with sugar-water and experienced sore throat pain,
had they used licorice instead,
more than half of them would likely have avoided the pain at time point $1.5$.
This result shows the effectiveness of licorice gargling in reducing sore throat pain from the causal attribution perspective and reinforces the previous conclusion.

\begin{table}[H]
\centering
\caption{Bound estimates and $95\%$ confidence intervals for $\PN$ at different time points}\label{tbl:app}
\begin{tabular}{ccccccc}
\hline
\multirow{2}{*}{Outcome}  & \multicolumn{2}{c}{Bound estimates} & \multicolumn{4}{c}{Confidence intervals} \\
& $[L_{\PT}, U_{\PT}]$ & $[L,U]$ & $\CI_1$ & $\CI_1'$ & $\CI_2$ & $\CI_3$ \\
\hline
$Y^{(1)}$   & $[0.481, 1]$ &  $[0.501, 1]$ & $[0.248,0.713]$ & $[0.165, 0.798]$ & $[0.256,1]$ & $[0.295,1]$\\
$Y^{(2)}$   & $[0.710, 1]$ &  $[0.732, 1]$ & $[0.539,0.881]$ & $[0.422, 1]$ & $[0.548,1]$ & $[0.578,1]$ \\
$Y^{(3)}$   & $[0.542, 1]$ &  $[0.583, 1]$ & $[0.355,0.730]$ & $[ 0.303, 0.829]$ & $[0.391,1]$ & $[0.422,1]$ \\
\hline
\end{tabular}
\end{table}

\section{Discussion}\label{sec:dis}

We establish a framework for statistical inference about the 
causal attribution measure $\PN$ under partial identification,
including efficient estimation of the covariate-assisted bounds and confidence intervals for $\PN$.
The proposed approach relies on the margin condition for estimating non-smooth bound functionals;
when this condition is implausible, the results should be interpreted with caution.
In such cases, one may consider using coarser covariate strata or estimating smooth approximations of the bounds \citep{levis2025covariate}.
Although our focus is the inference for $\PN$ in randomized experiments, 
the approach can be extended to observational studies.
For instance,
based on bounds for $\PN$ under ignorability \citep{kuroki2011statistical},
one can analogously construct bound estimators and confidence intervals.
More broadly, once bounds are established under suitable assumptions, similar inferential procedures can be applied to other causal attribution measures \citep{pearl1999probabilities,yamamoto2012understanding,lu2023evaluating, li2024retrospective, luo2025causal,rubinstein2025mediated}.
It is of interest to pursue these extensions in future research.

\section*{Supplementary materials}
\label{SM}

Web Appendices available online include
one-sided confidence intervals for hypothesis testing,
discussion of the testability of the margin condition,
a confidence interval based on the no-covariate bound,
proof of theorems and propositions,
and additional simulation and real data analysis results.

\bibliographystyle{apalike} 
\bibliography{CausalMissing.bib}
\end{document}